# Frequency-Modulation Mode-Locked Laser with GHz Spectral Width Tunable in the 2-3 µm Region

ZHEYUAN ZHANG [1], XIANGBAO BU [1], DAIKI OKAZAKI [2],
WENQING SONG [1], IKKI MORICHIKA [1], SATOSHI ASHIHARA [1, *]

[1]*Institute of Industrial Science, The University of Tokyo, 4-6-1 Komaba Meguro-ku, Tokyo 153-8505, Japan*
[2]*Institute for Chemical Research, Kyoto University, Ikenoura Gokasho, Uji, Kyoto 611-0011, Japan*
*\*ashihara@iis.u-tokyo.ac.jp*

**Abstract:** A narrow-bandwidth actively mode-locked laser using a Cr:ZnS gain medium has been successfully demonstrated. A free-space electro-optic phase modulator is employed in the solid-state laser resonator to achieve frequency-modulation (FM) mode-locking, which achieves a narrow spectral width of ~1 GHz and a pulse duration of ~500 ps over a wide tuning range of 1947-2445 nm. The operation frequency of the modulator determines the repetition rate of the mode-locked pulse train and can stabilize it to millihertz-level without any additional feedback loop systems. We also study the theoretical expression of pulse duration and spectral width in a FM mode-locking in a laser cavity that contains considerable group-delay dispersion. The results indicates that larger intracavity dispersion can only stabilize the laser operation by avoiding mode switching, but also narrow the spectral width and increase the pulse duration. The proposed laser features a narrow spectral width at a desired mid-infrared wavelength and a comb-like spectral structure with stabilized longitudinal mode spacing, providing a powerful tool for sensing and control of molecules.

## 1. Introduction

Mode-locked lasers capable of generating ultrashort pulse trains have advantages in many aspects, including high peak power, low shot-to-shot noise, and superior temporal/spatial coherence. To date, the development of mode-locked lasers has been primarily focused on broad spectral bandwidth and short duration. On the other hand, narrow-bandwidth mode-locked tunable lasers have attracted much interest in recent years due to their high power spectral density (PSD), arbitrary wavelength accessibility, and moderate peak intensity resulting from the relatively long pulse duration. These merits make the narrow-bandwidth mode-locked tunable lasers attractive for achieving efficient light-matter interactions in the applications of quantum optics [1], quantum information processing [2, 3], coherent control [4], molecular physics [5], and photoacoustic spectroscopy [6], etc. Tunable mode-locked lasers with a sub-nanometer spectral width of less than 10 GHz have been demonstrated in the near-infrared (NIR) region below 2 µm [7-11], however, few have been reported in the mid-infrared (MIR) region above 2 µm, where many vibrational resonances exist.

Although some progress has been made in the development of mode-locked tunable lasers operating in the 2-3 µm region, the spectral width is generally insufficiently narrow. Compared to 1-2 µm lasers, lasers in the 2-3 µm region have high efficiency when converted to longer wavelengths (5-8 µm). In 2016, W. B. Cho, et al. demonstrated a graphene mode-locked tunable Cr:ZnS laser, where a prism pair and a knife edge are used as a bandpass filter for wavelength tuning [12]. Although the tunable range was as wide as ~300 nm around 2.3 µm, the spectral width was 9.7 nm (~550 GHz) because of the femtosecond order pulse duration. In 2017, Y. Shen, et al. developed a narrow-bandwidth tunable Er:ZBLAN fiber laser, which was mode-locked by a semiconductor saturable absorber mirror (SESAM) [13]. This work achieved a spectral width of 1.5 nm (~59 GHz), but the tunable range was restricted to 2710-2820 nm due to the gain bandwidth of Er:ZBLAN fiber. In 2020, another work using Er:ZBLAN fiber was demonstrated by M. Pawliszewska, et al., where a 0.27 nm (9.6 GHz) spectral width and a picosecond pulse duration were realized in the range of 2840-2940 nm [14]. However, it is important to note that this narrow-bandwidth pulsed oscillation was achieved using the frequency-shifted feedback method, which is not considered a mode-locking technique because the longitudinal modes are not phase-locked [15].

Whilst many of the approaches described earlier used passive mode-locking techniques, active mode-locking techniques offer another promising way to achieve narrow-band mode-locking. Unlike passive mode-locking which relies on fast nonlinear effects (e.g., saturable absorption, Kerr effect), active mode-locking generates optical pulses with a narrower spectral width and longer temporal duration due to the slower active modulation. In the NIR region, active mode-locking is typically achieved through amplitude modulation (AM), where usually a waveguide-type intensity modulator is employed in the resonator to provide a periodic loss at a rate that is a multiple integer of the free spectral range (FSR) of the cavity [16]. For solid-state lasers, the cavity length is typically several meters, requiring a modulation frequency of at least tens of megahertz. However, few free-space intensity modulators above 2 µm can reach the required modulation frequency. Alternatively, phase modulators up to 20 GHz are commercially available in the MIR region, allowing the use of frequency modulation (FM) mode-locking in this region. FM mode-locking also has an advantage over AM mode-locking in that it does not require a biasing voltage, thus avoiding DC bias drift [17, 18].

In this paper, we demonstrate a narrow-bandwidth FM mode-locked Cr:ZnS laser that is broadly tunable in a 500 nm range near 2.3 µm. Cr:ZnS or Cr:ZnSe gain media represent exceptionally broad emission spectra in the 2-3 µm range [19, 20] and are therefore attractive for widely tunable lasers as well as for femtosecond pulse generation [21-27]. By employing a free-space electro-optic (EO) phase modulator and a simple grating-pair-slit configuration in a solid-state resonator, we realize a narrow spectral width of ~1 GHz and a ~500 ps pulse duration within a wide tunable range from 1947 nm to 2445 nm. To the best of our knowledge, this is the first FM mode-locked laser demonstrated at the wavelengths longer than 2 µm and achieves the narrowest spectral width in this region. The proposed laser with comb-like spectral structure also possesses a fixed mode spacing with mHz-level stability. It is expected to be useful for efficient excitation of molecular vibrations, providing powerful tools for applications that require high-resolution spectroscopy and control of molecules and organic materials.

## 2. Theory of FM mode-locking

In FM mode-locking, the optical spectrum of the light in the cavity is broadened by a phase modulator and constrained by a band-pass filter (or gain bandwidth) [28]. When the modulation frequency is identical to the free spectral range (FSR) of the laser resonator, the optical frequency of the light in the cavity will be periodically upshifted and downshifted, finally suppressed by the band-pass filter, leaving only the light that passes through the modulator without frequency shift will sustain and forms a mode-locked pulse train.

The theory of FM mode-locking has been intensively investigated in the last decades. For a phase modulator with a modulation frequency of $\omega_m$, the transmission function can be found in the work of Kuizenga and Siegman as $\exp(-j\Delta_m \cos \omega_m t)$ [29], where $\Delta_m$ is the maximum single-pass phase retardation. Based on the analyses in [17], the Haus master equation for an FM mode-locked laser in a linear cavity with negligible optical nonlinear effect can be expressed as

$$\left[g - l + \frac{4g_0}{\omega_G^2}\frac{d^2}{dt^2} + jD\frac{d^2}{dt^2} - S_m j\Delta_m \omega_m^2 t^2 + j\varphi\right] a(t) = 0, \qquad (1)$$

where $g$ and $l$ donate the gain and loss of the cavity, respectively, $g_0$ is the small signal gain, $\omega_G$ is the full width at half maximum (FWHM) of the gain line, $D$ is the total intracavity group delay dispersion (GDD) in the unit of s², $S_m$ represents the direction of phase modulation, $\varphi$ is the phase delay pre roundtrip, and $a$ express the temporal distribution of the mode-locked pulse. $S_m$ can be either positive or negative, corresponding to each extremum of the phase variation ($\cos \omega_m t$ equals 1 or -1), respectively. The two expressions describe the two situations in which the pulse train operates in either an up-chirped or a down-chirped mode. In a dispersion-free cavity, these two operating modes can get equivalent gain so the laser output may switch between them randomly, causing problems in laser stability. Introducing dispersion to the cavity can stabilize the operation mode by compressing one of the pulse trains whilst stretching the other one [17, 30].

After removing the time-independent terms in the master equation, it has a similar form to the time-independent Schrödinger equation, so its solution is Gaussian $a(t) = A_0 \exp(-t^2/2\tau_c^2)$. By substituting the solution back into the master equation, we obtain

$$\tau_c^2 = \sqrt{\frac{|D|}{2\Delta_m \omega_m^2 \gamma}} (A - jS_m B) \tag{2}$$

with $A$ and $B$ denoted as

$$A = \sqrt{\sqrt{\gamma^2 + 1} + S_m S_D \gamma}, \quad B = \sqrt{\sqrt{\gamma^2 + 1} - S_m S_D \gamma},$$

where $\gamma = |D|\omega_G^2/4g_0$, and $S_D$ represent the signs of $D$. The pulse duration can then be estimated as

$$\Delta t = \frac{1}{\pi} \left(\frac{2 \ln 2}{f_m f_G}\right)^{\frac{1}{2}} \left(\frac{2g_0}{\Delta_m}\right)^{\frac{1}{4}} \left[(\gamma^2 + 1)\left(\sqrt{\gamma^2 + 1} - S_m S_D \gamma\right)\right]^{\frac{1}{4}}, \tag{3}$$

where $f_m = \omega_m/2\pi$ and $f_G = \omega_G/2\pi$. Through the Fourier transformation, the mode-locked pulse in the frequency domain can be expressed as $A(f) = A_0\sqrt{2\pi\tau_c^2}\exp[-2\tau_c^2\pi^2(f - f_c)^2]$, where $f_c$ is the central frequency, so the spectral width can be similarly acquired as

$$\Delta f = 2(\ln 2\, f_m f_G)^{\frac{1}{2}} \left(\frac{\Delta_m}{2g_0}\right)^{\frac{1}{4}} \left(\sqrt{\gamma^2 + 1} - S_m S_D \gamma\right)^{\frac{1}{4}}. \tag{4}$$

The expression of spectral width also includes an additional term that is dependent on the value of $\gamma$. When the dispersion is negligible, $\gamma$ is regarded as 0, so the pulse duration and spectral width can be obtained as $\Delta t_{D=0} = 1/\pi \cdot (2 \ln 2 /f_m f_G)^{1/2} (2g_0/\Delta_m)^{1/4}$ and $\Delta f_{D=0} = 2(\ln 2\, f_m f_G)^{1/2}(\Delta_m/2g_0)^{1/4}$, which are identical to what is predicted in Ref. 27.

For a specific laser, the sign of dispersion $S_D$ depends on whether the GDD is positive or negative, whilst $S_m$ alternates between positive and negative during the modulation, resulting in two pulse trains with opposite chirp directions coexisting in the cavity. When $S_m S_D$ is negative, the pulse train will have a broader spectral width as understood from eq. (4), and therefore, suffer from larger loss due to the limited gain bandwidth. As a result, $S_m S_D$ can always be considered positive for the mode-locked pulse train.

Deriving Eq. 3 and 4 with respect to $\gamma$ reveals that the pulse duration increases with larger $\gamma$ values when the dispersion is significant, whilst the spectral width decreases monotonically as $\gamma$ increases. These findings suggest that a low modulation frequency, small modulation depth, and large dispersion are favorable for achieving a narrow spectral width. Experiments have demonstrated the relationship between spectral width and modulation depth [31, 32]. It has also been reported that pulse duration has a positive correlation with GDD [17], so it is reasonable to expect that dispersion can play a role in narrowing the spectral width. This can be understood in a way that the mode-locked pulses with compensated chirps need fewer frequency components to form a stable reproducible pulse train.

## 3. Experimental setup

The setup of the FM mode-locked Cr:ZnS laser is shown in Fig. 1a. The laser is a linear cavity with a total length of ~1.5 m, which corresponds to a free spectral range (FSR) of ~100 MHz. The pump source used in this experiment is an erbium-doped fiber laser (EDFL, Keopsys,

CEFL-KILO-10-LP) that emits a linearly polarized continuous wave at 1565 nm with a kHz-level linewidth. The pump light is focused on the Cr:ZnS polycrystal through a concave dichroic mirror using a convex lens. The Cr:ZnS polycrystal, which has AR coating on its both sides, has a length of 4.8 mm and a Cr concentration of $5.0\times10^{18}$ cm$^{-3}$. The mode-locked output power is typically ~100 mW at a pump power of 2 W.

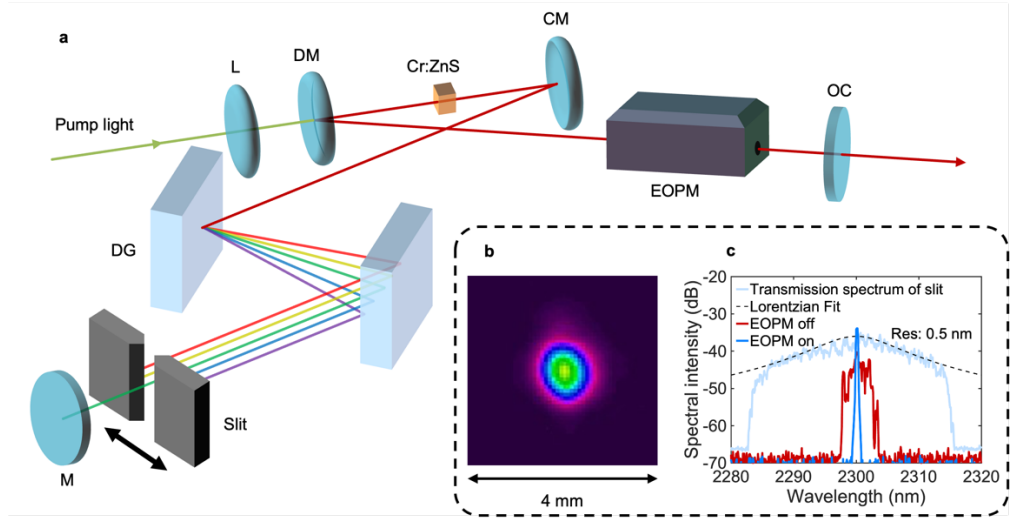

Fig. 1. (a) Setup of the FM mode-locked laser. EDFL: erbium-doped fiber laser, L: plano-convex lens, DM: dichroic concave mirror, CM: concave mirror, DG: reflective diffraction grating, HR: highly reflective mirror, EOPM: electro-optic phase modulator, OC: output coupler. (b) Beam profile of the laser output. (c) Transmission spectrum of the bandpass slit (light blue) and its Lorentzian fitting (dashed), and spectrum of the laser output when the modulator is off (red) and on (blue).

A pair of Au-coated reflective gratings with a line density of 600 lines/mm (Spectrogon, PC-0600-50x50x10-NIR) is used to provide dispersion. As mentioned in the last section, the existence of dispersion can stabilize the laser by avoiding the mode switching. It can also broaden the resonances of the laser cavity and lower the threshold for frequency matching [33]. The angle of incidence to the first grating is ~65° and the spacing between the two gratings is 80 mm, which results in a total anomalous dispersion of approximately -1.56 ps$^2$ (0.588 ps/nm, @ 2200 nm). The Cr:ZnS polycrystal has a GDD of approximately +600 fs$^2$, whilst the EO phase modulator has a GDD of +1000 fs$^2$ @ 2.0 μm and +8000 fs$^2$ @ 2.5 μm. As the GDD of all other components is negligible, the dispersion provided by the grating pair is dominant in the total intracavity dispersion. A slit is placed in front of the highly reflective mirror to provide the bandpass filtering required for FM mode-locking. To prevent additional loss, the size of the slit is set to 1.5 mm. Due to the large dispersion provided by the grating pair, the exact cavity length and FSR vary with the wavelength. Therefore, the laser will only be mode-locked when the modulation frequency of the phase modulator matches the FSR of the wavelength selected by the slit.

The EO phase modulator (Qubig, PM8-SWIR2_100) is positioned near the output coupler to minimize the phase walk-off between the light traveling in two directions. The lithium niobate (LN) crystal used for modulation has a length of ~3 cm and an AR coating for the wavelength range of 1.7-3.0 μm. Its modulation frequency can be tuned between 83.7 and 117.0 MHz. Empirical data suggests that the detuning between the modulation frequency and the FSR should be less than ~10 kHz to achieve stable mode-locking. The beam profile of the laser output during mode-locking measured by a beam profiler is shown in Fig. 1b, showing a beam with Gaussian distribution and a beam size of ~1 mm.

The transmission spectrum of the slit functioning as the bandpass filter is measured by an optical spectrum analyzer (OSA, Yokogawa Test & Measurement Corp., AQ6375E, 1000-2500 nm) and shown in Fig. 1c as the light blue trace. It is measured during CW operation by partially blocking the slit from side to side and recording the max-hold spectrum of the laser output. The transmission spectrum has a Lorentzian distribution within a range of ~30 nm and a 3 dB width of ~12 nm. The output spectrum of the laser during continuous wave (CW) operation and FM mode-locking are also shown in Fig. 1c. The CW operation spectrum exhibits an unstable distribution of lasing at random wavelengths in a range of ~ 5 nm with random amplitudes due to the intense competition between longitudinal modes. At a modulation frequency of 100.4 MHz, the laser is mode-locked with a single spectral peak centered at 2300 nm and a spectral width narrower than the resolution of 0.5 nm. The reason for the spectra narrowing is that the gain is saturated by the mode-locked spectral peak, which prevents the other longitudinal modes from lasing in CW operation mode due to the cross-saturation effects.

## 4. Results

The performance of the laser is shown in Fig.2. Fig. 2a displays the oscilloscope trace with 32 times of averaging when the laser is mode-locked at 2200 nm with a modulation frequency of 101 MHz. The pulse spacing is measured to be 10 ns, which is approximately identical to the inverse of the modulation frequency. Slight increases in optical power can be observed between the neighboring pulses, which originates from the frequency-modulated light that has opposite phase retardation with the dominant mode-locked pulses. These minor pulses with opposite direction of chirp cannot get enough gain due to the broader spectral width and thus larger loss, so can be disregarded in most situations. The autocorrelation trace was measured using a homemade autocorrelator with an 80 cm moving range of the delay line after filtering the possible pump (1565 nm) and the second harmonic components (1015 nm) with long-wavelength-pass filters. Fig. 2b shows the measured autocorrelation trace of the mode-locked pulse, which has a width of 707 ns. Since the autocorrelation trace is Gaussian, the pulse duration can be obtained as 500 ps in FWHM using the deconvolution factor of $\sqrt{2}$.

The spectral width measured by a scanning Fabry-Perot interferometer (Thorlabs Inc., SA210-18C, 1800-2600 nm, FSR 10 GHz, resolution 67 MHz) is shown in Fig. 2c. It has a narrow spectral width of 1.2 GHz in FWHM at 2200 nm and fits well with a Gaussian distribution. The comb-like spectral structure with only 12 longitudinal modes within the FWHM can be observed in the inset of Fig. 2c. By multiplying the pulse duration and the spectral width, the TBP of the FM mode-locked laser can be obtained as 0.6. By adjusting the RF power of the modulator driver, we can change the modulation depth. Fig. 2d shows the spectral width as a function of the single-pass modulation depth, which shows a positive correlation as Eq. 4 predicted.

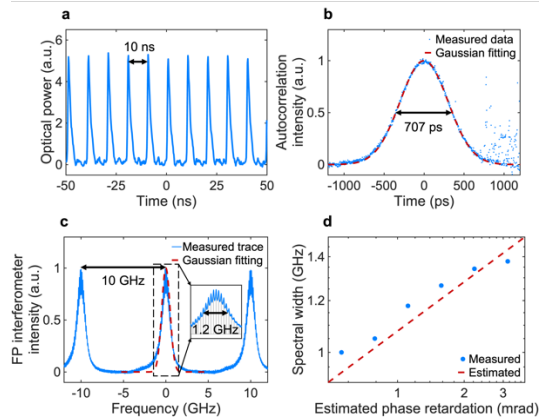

**Fig. 2.** (a) Oscilloscope trace, (b) intensity autocorrelation trace, and (c) FP interferometer trace of the laser output when it is mode-locked at 2200 nm with a modulation frequency of 101 MHz. (d) Measured spectral width as a function of the estimated phase retardation.

By changing the location of the slit and correspondingly tuning the modulation frequency of the phase modulator, the central wavelength can be tuned from 1947 nm to 2445 nm. Fig. 3a shows the corresponding slit position and modulation frequency as a function of the central lasing wavelength. The FSR, or the modulation frequency, which is equivalent to the repetition rate, is the inverse of the cavity round-trip time, so $D$ can be obtained from the slope of round-trip time as a function of lasing wavelength $d(1/\text{FSR})/d\lambda$, resulting in a value of 0.42-0.83 ps/nm for $D$ (wavelength-dependent). The value of $\gamma$ can then be taken as 26-54. The spectra measured with these parameters are shown in Fig. 3b, which shows a wide tuning range of ~500 nm and good spectral flatness. The light blue line represents the envelope that is measured during CW operation in the max-hold mode, which has similar distribution with the envelope of the mode-locked spectrums. The spectral intensity is lower near 2500 nm due to the water absorption in the air. Note that the tuning of the lasing wavelength is not perfectly continuous, an unequally spaced tuning step of 0.1-0.7 nm exists during the wavelength tunning. It may originate from the wavelength-dependent loss in the optical components and can be reduced by lowering the bandwidth of the bandpass filter.

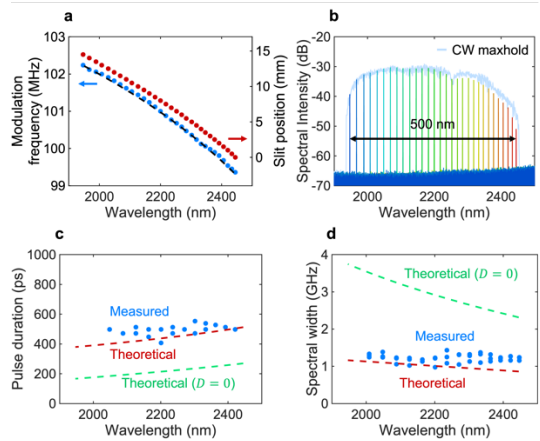

**Fig. 3.** (a) Corresponding slit position (blue) and modulation frequency (red) for each lasing wavelength. (b) Optical spectrum of the laser output in the tunable range. (c) Pulse duration and (d) spectral width of the mode-locked pulses as a function of the lasing wavelength. Blue: measured, red: theoretical values with large GDD, green: theoretical values with negligible GDD.

Fig. 3c and 3d show the measured pulse duration and spectral width, respectively, as a function of the lasing wavelength. The estimated value of $\Delta t$ and $\Delta t_{D=0}$ are shown in Fig. 3c, whilst $\Delta f$ and $\Delta f_{D=0}$ are shown in Fig. 3d. As illustrated in Fig. 1c, the 3 dB bandwidth of the bandpass filter is ~12 nm, allowing for an estimation of the gain bandwidth $f_G$ to between 0.46-1.17 THz (depending on the wavelength). The measured results agree well with the theoretical values indicated by eq. (3) and (4) but show no discernible wavelength dependency due to the moderate measurement accuracy. The quantitative deviation between the measured data and the theoretical values may originate from the nonlinear effect in the Cr:ZnS crystal or the possible wavelength dependence of phase modulation.

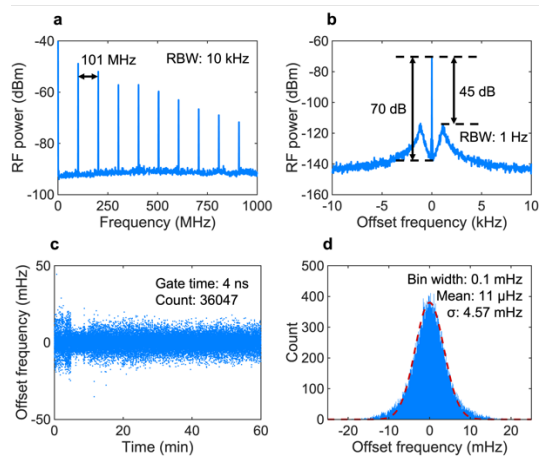

**Fig. 4.** RF spectrum of the laser operated at a modulation frequency of 101 MHz in the span of (a) 0-1 GHz and (b) 20 kHz centered at 101 MHz. (c) Frequency deviation between the laser repetition rate and the modulation frequency over a period of one hour. (d) Histogram of the frequency deviation.

The RF spectrum of the laser output is shown in Fig. 4. Fig. 4a shows the RF spectrum in the 0-1 GHz range. The frequency of the fundamental order and the frequency spacing between higher harmonic orders are uniformed and identical to the modulation frequency, which demonstrates that the repetition rate of the mode-locked pulse train is determined by the modulation frequency. Fig. 4b shows a measurement taken within a span of 20 kHz near the 101 MHz peak, which illustrates a 3 dB bandwidth of 0.9 dB when the RBW is 1 Hz. Two sub-peaks 45 dB lower are observed ~1 kHz away from the main peak, which originate from the servo bumps of the driver of the modulator, so is expected to be suppressed by replacing the driver with more advanced electronic devices in the future. With a suppressed servo bump, the SNR is expected to be 70 dB. Fig. 4c illustrates the measured frequency deviation between the laser repetition rate and the modulation frequency over a period of one hour. The frequency deviation is measured using a frequency counter (Keysight Technologies, 53220A). The laser output is converted into an RF signal by an HgCdTe (MCT) photovoltaic detector (Vigo photonic, PEM-10.6). The signal is input into the frequency counter with an electronic bandpass filter (Mini-Circuits, BBP-101+, 94 - 108 MHz) and an RF amplifier. A reference signal from the modulator driver is split and directly input into the other channel of the frequency counter so that the frequency deviation can be acquired. The gate time is set to 4 ns, and 36047 points were recorded. The histogram of the frequency deviation with a bin width of 0.1 mHz is shown in Fig. 4d, which has a normal distribution and illustrates a standard deviation of only 4.57 mHz. These results demonstrate that the phase modulator can precisely stabilize the repetition rate to the modulation frequency without requiring any additional feedback loop system.

## 5. Discussion

In this research, we study the theoretical expression for pulse duration and spectral width of FM mode-locking under considerable intracavity dispersion. The derived equations indicate that the existence of intracavity dispersion can reduce the spectral width of an FM mode-locked laser. We also for the first time demonstrate an FM mode-locked Cr:ZnS laser, which achieves a narrow spectral width of ~1 GHz in the range from 1947 nm to 2445 nm. Whilst most mode-locked MIR lasers typically rely only on a bandpass filter for wavelength narrowing and fail to reach a narrow spectral width at the level of the sub-nanometer, the FM mode-locking technique with low modulation depth and large intracavity dispersion can significantly reduce the spectral width to the GHz level. The proposed narrow spectral width mode-locked laser in the fingerprint region can be considered as a super narrow frequency comb with a millihertz-level-stabilized repetition rate and extended to MIR region >3 μm through the optical parametric process, which is expected to be useful for many applications, especially for the spectroscopy and molecule physics. Moreover, due to the absence of a high-speed intensity modulator at the MIR region, we also expect the FM mode-locking to be a simple solution for high repetition rate mode-locking in the MIR region where the short cavity is not practical.

**Funding.** This work is supported by Japan Society for the Promotion of Science (20H02651, 20J22067, 20K20556, 21K14584), Core Research for Evolutional Science and Technology (JP20348765), and Ministry of Education, Culture, Sports, Science and Technology (Q-LEAP JPMXS0118068681).

**Disclosures.** The authors declare no conflicts of interest.

**Data availability.** Data underlying the results presented in this paper are not publicly available at this time but may be obtained from the authors upon reasonable request.